\documentclass{ifacconf_amx}

\usepackage{graphicx}      
\usepackage{amsmath,amssymb,amsfonts} 
\usepackage{xcolor}        
\usepackage{cite}          

\usepackage{siunitx}       
\usepackage{fancyhdr}      

\usepackage{lipsum}        

\usepackage[scaled]{helvet}
\usepackage[T1]{fontenc}

\newcommand{\ISBN}{UNKNOWN ISBN}

\newcommand{\PaperTitle}{A title for your paper}
\newcommand{\PaperDate}{01 April 1900}
\newcommand{\AuthorFooter}{Oksanen T.}

\makeatletter
\let\old@ssect\@ssect 
\makeatother

\usepackage[hidelinks]{hyperref}      
\usepackage{cleveref}      

\makeatletter
\def\@ssect#1#2#3#4#5#6{%
	\NR@gettitle{#6}
	\old@ssect{#1}{#2}{#3}{#4}{#5}{#6}
}
\makeatother

\usepackage{tikz}
\usepackage{stfloats}

\renewcommand{\ISBN}{978-3-911430-07-4}
\renewcommand{\PaperTitle}{ISO FastLane: Faster ISO 11783 with Dual Stack Approach as a Short Term Solution}
\renewcommand{\PaperDate}{10 Feb 2026}
\renewcommand{\AuthorFooter}{Oksanen Timo}

\hypersetup{
  pdftitle   = {ISO FastLane: Faster ISO 11783 with Dual Stack Approach as a Short Term Solution},
  pdfauthor  = {Timo Oksanen},
  pdfkeywords= {SAE J1939, CAN bus, ISO 11783, High Speed ISOBUS, protocol extension, in-vehicle networks, standardization},
  pdfcreator = {LaTeX}
}

\begin{document}
\begin{frontmatter}

   \title{\PaperTitle}

   \author{Timo Oksanen}

   \address[First]{Technical University of Munich (TUM), Germany; Professorship of Agrimechatronics, Germany; e-mail: timo.oksanen@tum.de}
   \address[Second]{Munich Institute of Robotics and Machine Intelligence (MIRMI), Germany}

   \address[Third]{Aalto University; Department of Electrical Engineering and Automation, Finland}

   \begin{abstract}
      The agricultural industry has been searching for a high-speed successor to the 250~kbit/s CAN bus backbone of ISO~11783 (ISOBUS) for over a decade, yet no protocol-level solution has reached standardization. Meanwhile, modern planters, sprayers, and Virtual Terminals are already constrained by the bus bandwidth. This paper presents ISO FastLane, a gateway-less dual-stack approach that routes point-to-point ISOBUS traffic over Ethernet while keeping broadcast messages on the existing CAN bus. The solution requires no new state machines, no middleware, and no changes to application layer code: only a simple Layer~3 routing decision and a lightweight peer discovery mechanism called Augmented Address Claim (AACL). Legacy devices continue to operate unmodified and unaware of FastLane traffic. Preliminary tests reported on the paper demonstrate that ISO FastLane accelerates Virtual Terminal object pool uploads by factor of 8 and sustains Task Controller message rates over 100 times beyond the current specification limit. Because ISO FastLane builds entirely on existing J1939 and ISO~11783 conventions, it can be implemented by ISOBUS engineers in a matter of weeks. This is delivering tangible performance gains today, without waiting for the long-term High Speed ISOBUS solution.
   \end{abstract}


\end{frontmatter}

\section{Introduction}
\label{sec:introduction}

The ISO 11783 standard series (market name ISOBUS) was started in 1991, when ISO TC 23/SC 19 and its WG1 were founded \cite{Oksanen2021}. The decision to use CAN bus \cite{ISO11898-1} as a backbone for this network was evident, as it was already technology of DIN 9684 series, decided in 1988, and also for SAE J1939 \cite{J1939}. In the early 1990's, 250 kbit/s (kbaud) was a good compromise between the imagined functions  and physical layer limitations of CAN technology. However, over the years, agricultural machines have become more and more sophisticated in mechatronics and sensors.

For the modern tractor-implement system, 250 kbit/s has already been a limiting factor for data throughput. For instance, high-tech planters have row specific section control and rate control and the operating speed has increased dramatically. The multi-rate commands and section control would exceed the allowed capacity of the ISO 11783 bus load in ISO 11783-10 \cite{ISO11783-10} protocol. Also, nozzle based section control for wide and fast sprayers was beyond imagination in the 1980's. Also, the virtual terminal update rates and pool upload with objects with true colours are limiting both innovations and user experience.

ISO TC 23/SC 19/WG1 started a task force to study future wired communication backbones beyond CAN bus already in 28th March 2012 (Jacob van Bergeijk, David Smart, Marco Ferretti, Giorgio Malaguti, Martijn van der Bijl and Timo Oksanen). The work has continued under AEF since 2014 to explore technologies beyond CAN bus for a future backbone. This work to find the long-term solution is still ongoing. The exploration is extended now from agriculture to other vehicles and there is Joint Working Group 16 (ISO TC 127/SC 3/JWG 16) to find a common solution. The current achievement of the AEF project team is the physical layer solution with dedicated connector for ``High Speed Isobus (HSI)'' and cabling requirements. However, the protocol selection is wide open at time of this publication writing.

Agricultural electronics has success stories. ISOBUS itself is a success story. No other vehicle or automation industry has been able to agree on a common technology where no competing technologies exist worldwide. All manufacturers are committed on the same technology.

The first success story was already in 1980's, when the standardization took first steps. In 1986, DIN committee made a decision to split the required standardization to two parallel projects: short-term solution and long-term solution. The short-term solution was needed very soon and this allowed more time to complete research for a long-term solution. The short-term solution for tractor-implement communication was the DIN 9684-1 ``signal connector'' which was converted later to ISO 11786 \cite{ISO11786} standard. This standard was started in 1986, the draft was ready in 1987 and the final published standard was 1989. This ``signal connector'' is still available as an option for some new tractors rolling out from tractor factories. Thus, we can easily say that the short-term solution has at least a lifetime of 50 years in the farms. The long-term solution was started in 1987 and this led to bus system architecture and DIN standard 9684 parts 2-5, which were known as ``LBS''. The current solution ISO 11783 is actually the same continuum of long-term solution as the basic principles and technology are still the same. Short-term solution development took 1 year, long-term solution development took at least 15 years, before all required parts of ISO 11783 were done around 2004.

The High-Speed Isobus project needs similar approach: a short-term solution and a long-term solution, which need to be developed in parallel. The short-term solution needs to be as simple as possible, so that development is complete in maximum one year. Known long-term solution proposals include OPC UA \cite{Siponen2022} and DDS \cite{Brodie2024}. 

In this paper, we present one candidate for the short-term solution. This solution utilizes both CAN bus and Ethernet in parallel for those devices which have Ethernet connectivity. Legacy devices work as earlier and they do not even notice the existence of ISO FastLane devices which talk to each other. The beauty of this proposal is simplicity, but every solution also has drawbacks. 

\section{Requirements}
\label{sec:requirements}

\begin{table*}[!b]
\caption{Requirements for short-term and long-term solution}
\label{tab:requirements}
\centering
\small
\begin{tabular}{|c|p{6.5cm}|p{4cm}|p{5cm}|}
\hline
\textbf{\#} & \textbf{Requirement} & \textbf{Short-term} & \textbf{Long-term} \\
\hline
1 & Physical CAN bus mandatory for each ECU HW having ISOBUS CF & Yes & No \\
\hline
2 & Efficiency optimization on Ethernet & No & Yes \\
\hline
3 & Service discovery mechanisms on Ethernet & No & Yes \\
\hline
4 & Discovery in practice & Provided by current ISO 11783 & Provided by HSI \\
\hline
5 & Modifications required to ISO 11783 stack & Minimal & Refined functionalities \\
\hline
6 & Network management complexity on Ethernet & Minimal & Full hot plug-and-play required \\
\hline
7 & Separation of data exchange and information model & No modification to current ISO 11783 & Modern architecture required \\
\hline
8 & Connector for tractor-implement & Both current IBBC + HSI connector mandatory & Only HSI connector mandatory \\
\hline
9 & Connector for in-cab devices & Both In-Cab and HSI connectors mandatory & Only HSI connector mandatory \\
\hline
10 & Requires further research & No & Yes \\
\hline
11 & Duration of guideline completion (target) & Few months & Few years \\
\hline
\end{tabular}
\end{table*}

\subsection{Short-term vs Long-term}
\label{subsec:shortvslong}

Typical for every engineering is to find a solution that fulfils the requirements. Sometimes emerging need for solution pushes time over other requirements, such as flexibility, optimality or cost. Common approach in these situations is to split the project into parallel branches, where one subproject aims to deliver a functional solution in short time even if that is not optimal and the other subproject has time to do further research to find the optimal solution. Typical for short-term solutions is that they are based on the best available idea without any further research and rely on existing technologies that can be implemented immediately.

In Table~\ref{tab:requirements}, an illustration for short-term and long-term requirements in this context are presented. Long-term solution has been prepared for years, so those requirements are obvious. Short-term solution requirements are derived from the principle of simplicity and minimal changes to existing systems.

\subsection{Backwards Compatibility with the ISOBUS System}
\label{subsec:backwards}

Backwards compatibility is very important requirement in agricultural electronics. The lifetime of electronics equals to lifetime of vehicle if electronics is embedded. Lifetime of tractor or implement is easily 20 years. Farms have mixed generations of electronics in use, either combinations of old tractors with new implements and other devices in the cab, and vice versa. Due to this reason, also mechanical interface standards between tractor and implement are difficult to modify (e.g. ISO 730, ISO 500, ISO 6489 et cetera).

In the short-term solution backwards compatibility means especially the requirement that legacy ECU's and their software realizing Control Functions (CF) can work as is, and no updates are needed for those. New designs may contain additional features, but they also need to operate with legacy devices. This can be realized either having two operating modes out of which the appropriate is selected after initial handshake, or there is only one operating mode but extended features are only used between CF's who are aware of those.

\subsection{Discovery}
\label{subsec:discovery}

Communication systems need to have discovery mechanism to find each other on the network after power-up. In static systems this might be even based on fixed addresses, but in ISO 11783 context this is impossible due to dynamic hot plug-and-play requirement. ISO 11783 \cite{ISO11783-5} and SAE J1939 \cite{J1939-81} have resolved the discovery by two means: a) address claiming which provides dynamic addresses but also information about device class and function of CF, and b) application layer specific mechanism. Examples of application layer specific mechanisms are VT status message which is broadcasted by VT server and similarly TC status message, which is also broadcasted by TC server. These allow every client to discover servers. Thus, the discovery mechanism is already existing in the current ISO 11783.

For the future comprehensive High Speed ISOBUS, it is required to provide modern discovery mechanism, to find services in common way, not coupled with addressing or with application layer. Internet technologies offer existing technologies for this. 

\subsection{Heartbeat}
\label{subsec:heartbeat}

For session connection monitoring, a mechanism is required to monitor that connection is still alive after initial handshake. In ISO 11783 context, this mechanism is known as heartbeat, which is implemented both in VT, TC and WSM (working set master) status messages (once in one or two seconds). Connection aliveness is monitored with timeout, if this status messages are not received any more, the defined shut-down and restart procedure shall be followed. Dedicated heartbeat defined in ISO 11783-7 \cite{ISO11783-7} is intended for other signals beyond VT and TC; such as speed signal originating from Tractor ECU (TECU).

Both short-term and long-term solution shall provide session monitoring. Typical middleware options for long-term solution provide this built-in. In short-term solution, we may use the existing ISO 11783 mechanism, but avoiding complex rules such as separate monitoring on CAN and High-Speed ISOBUS. Rules need to be simple. 

\subsection{Point-to-point Communication}
\label{subsec:pointtopoint}

Several functionalities on ISOBUS use point-to-point communication following client-server architecture in current ISO 11783. These utilize SAE J1939 PDU Type1 messages, which are destination specific messages. That is the only way to implement point-to-point communication on J1939 context. This communication pattern is used extensively in Virtual Terminal (ISO 11783-6 \cite{ISO11783-6}), Task Controller (ISO 11783-10 \cite{ISO11783-10}) and File Server (ISO 11783-13 \cite{ISO11783-13}) application layers. Virtual terminal (known as Universal Terminal, UT, by AEF functionality) is using PGN's called VT2ECU and ECU2VT, Task Controller is using PD messages and File Server is using FS messages.

Only exception in those protocols are the so called server-status messages, which are broadcast messages, for dual purpose: discovery and heartbeat. Discovery requires broadcast message type. All these point-to-point communication application layers are suitable for direct communication on Ethernet by using unicast messaging, such as UDP unicast. 

\subsection{Broadcast Communication}
\label{subsec:broadcast}

Another communication pattern on ISO 11783 is based on broadcast messages without subscription. These are mainly messages defined in ISO 11783-7 \cite{ISO11783-7}, transmitted by Tractor ECU. These messages contains signals such as vehicle speed, PTO speed, hitch position, engine speed, key switch state, lighting, and GNSS position. These are intended for all CF's who need that information and it is up to implementation of CF software if these are parsed for processing or filtered out.

Broadcast messages in ISO 11783 / SAE J1939 / NMEA 2000 \cite{NMEA2000} context are using either J1939 PDU Type2 messages, or PDU Type1 messages with global address as destination.

These broadcast messages are still core part of short-term solution as changing everything to point-to-point communication in service oriented architecture by small touches for ISO 11783 series is not possible.

These broadcast messages are more difficult to transmit on Ethernet, as even if Ethernet provides broadcast messaging, there are known problems with those in practice, in complex networks with managed switches. Broadcast messaging in the short-term solution needs to be as bullet-proof as today, without risks in the field. 

\subsection{Suitability for Existing ECU Hardware}
\label{subsec:related_standards}

Short-term solution is intended to be brought on the market in short-term. This means also that the solution needs to be such that it does not require extensive additional libraries, security chips, time-synchronized transceivers or other technologies that force industry to design completely new ECU hardware. These requirements would hinder bringing products on the market, as typical cycle times in ECU development for agricultural machinery are about five years. Naturally, the ECU's must have besides CAN interface the AEF HSI interface as required by those guidelines.

\section{The Fundamental Idea}
\label{sec:solution}


In this paper, we present the novel idea how to enable short-term solution with simple rules and respecting all requirements defined above. The fundamental ideas are:

\begin{enumerate}
\item[(a)] Every ECU must still have physical CAN bus interface as the primary interface:
    \begin{itemize}
    \item either built-in to ECU,
    \item or via proprietary gateway of same OEM.
    \end{itemize}

\item[(b)] Every ECU has additional HSI interface according to rules of AEF, as the secondary interface.

\item[(c)] Address claiming procedure in CAN bus is augmented with additional information about existence of HSI interface and its addressing:
    \begin{itemize}
    \item Augmented address claiming provides addressing information of their own IP addresses on secondary interface.
    \end{itemize}

\item[(d)] All broadcast messages (all PDU Type2 and PDU Type1 where destination address is global) are transmitted exclusively over CAN bus, as in the current ISO 11783 network:
    \begin{itemize}
    \item consequently address claiming happens in CAN bus exclusively, including augmented information sharing.
    \end{itemize}

\item[(e)] In point-to-point communication, when both server and client have identified that each other has secondary interface available with valid IP address, they send all point-to-point information to each other exclusively over secondary interface:
    \begin{itemize}
    \item removes all point-to-point VT2ECU and ECU2VT, PD and FS communication from physical CAN bus to Ethernet during that session.
    \end{itemize}

\item[(f)] Moving point-to-point communication from primary interface (CAN) to secondary (Ethernet) allows both reducing bus load for broadcast messaging allowing slight upgrade for broadcast messaging:
    \begin{itemize}
    \item This has both benefits for point-to-point protocols (VT, TC, FS), but also for broadcast (TECU).
    \end{itemize}

\item[(g)] Point-to-point communication over secondary interface (Ethernet) allows ramping up data exchange, number of messages per second, but keeping the same frames defined in ISO 11783, for instance:
    \begin{itemize}
    \item TC is limited now ``average 10 messages per second'' e.g. for section control. In secondary interface the rule could be turbocharged to 100 or even more.
    \item VT object pool upload using Extended Transport Protocol (ETP) happens in secondary interface.
    \item FS file transfer happens in secondary interface and this allows faster file transfer.
    \end{itemize}

\item[(h)] This idea provides a generic solution for OSI layer 3, which is application layer independent (works equally for VT, TC, FS etc):
    \begin{itemize}
    \item To turboboost the file transfer, object pool upload or fast section control, minimal exceptions are required for application layer rules that enable more messages per second than the current rules (e.g. 100 instead of 10).
    \end{itemize}

\item[(i)] Gateway-less design:
    \begin{itemize}
    \item Simplicity is key for short term solution success. Any gateway design required finite state machine designs, error handling, timeout management, etc. Design without gateway on secondary interface (Ethernet) simplifies the whole system and keeps the state management on CAN bus side completely. This allows easy extension of current software stacks to support ISO FastLane, as it requires minimum changes. 
    \end{itemize}
\end{enumerate}

\section{The Architecture}
\label{subsec:architecture}

This system requires that every Control Function in ECU has access at least to ISO 11783 network (CAN bus) and optionally to Ethernet. ISO FastLane requires both, but legacy devices are only in CAN bus. 

\subsection{Technical Solution: Indication of Secondary Channel}

As explained above, one of the fundamental ideas is to include indication of existence of secondary channel as part of augmented address claiming and details of that. The additional information needs to be communicated on CAN bus and this requires a new PGN for that purpose, which needs to be PDU Type1 destination specific. The details of this information contain at least IP address of secondary channel, either IPv4 or IPv6. IPv4 address is 32-bits and IPv6 address is 128-bits long. As this method aims at generic solution, with either address, the data length is at least 128-bits long, plus it requires at least one bit long header to indicate which address is the payload. Obviously, this leads to using Transport Protocol (TP) \cite{ISO11783-3} for moving this information over J1939 network.

As we want to provide augmented address claim, it makes sense to design it extendable, for other short-term future needs. Another need might come from Wireless In-Field communication project, as CF's could also indicate even further communication channels beyond primary and secondary. In addition, future diagnostics needs would benefit additional information. Therefore, the new PGN content should be scalable beyond IP address.

Our approach is to follow the same philosophy how VT and TC object pools are built. The binary serialization of objects is based on idea of standardized types, with fixed length. 

\begin{table}[htbp]
\caption{Objects for Augmented Address Claim (AACL) message}
\label{tab:aacl_objects}
\centering
\begin{tabular}{|c|l|l|l|}
\hline
\textbf{Object} & \textbf{Name} & \textbf{Attribute 1} & \textbf{Attribute 2} \\
\hline
1 & IPv4 address & 4 bytes: IPv4 address & 2 bytes: UDP port \\
\hline
2 & IPv6 address & 16 bytes: IPv6 address & 2 bytes: UDP port \\
\hline
3-255 & Reserved & - & - \\
\hline
\end{tabular}
\end{table}

\begin{table}[htbp]
\caption{AACL CAN Frame Structure (Prototype Values)}
\label{tab:aacl_frame}
\centering
\begin{tabular}{|l|l|}
\hline
\textbf{Field} & \textbf{Value (Prototype)} \\
\hline
PGN & 19200 (0x4B00) * \\
\hline
PF (PDU Format) & 75 (0x4B) * \\
\hline
Priority & 6 \\
\hline
29-bit CAN ID & 0x184BDA[SA] \\
\hline
\multicolumn{2}{|l|}{\small * Final PF to be assigned by SAE} \\
\hline
\end{tabular}
\end{table}

\textbf{Note:} Object 1 (IPv4) is 7 bytes total, fitting in single CAN frame without Transport Protocol. Object 2 (IPv6) is 19 bytes, requiring TP RTS/CTS. Both objects may be present (26 bytes total). The UDP port is per-CF; multiple CFs on same ECU sharing one IP address MUST use different ports. The recommended default port is 11783.

For prototyping purposes, the AACL message uses PGN 19200 (0x4B00, PF=75) with priority 6, matching the Address Claim message priority. This is a PDU Type 1 (destination-specific) message. Note: The final PF value shall be assigned by SAE through their formal standardization process; PF=75 is used here for prototype implementation only. The mechanism to trigger the sending this information in J1939 network is based on J1939 Request message in destination specific mode. The CF having the secondary interface shall always respond to the request.

A key design rule is that a CF shall have its IP address ready BEFORE claiming its J1939 address. This eliminates complexity of NULL value handling and retry logic. If a CF's IP address changes after initial claim, the CF shall perform a new address claim, which triggers all peers to re-request AACL.

The AACL request shall be sent after the standard 250ms address claim validity period (per ISO 11783-5 \cite{ISO11783-5} section 4.5.2). The response shall be sent back with destination-specific message. If no response is received within a reasonable timeout, the peer is assumed to not support ISO FastLane.

\subsection{Device Lifecycle: Entry, Presence and Departure}

ISO 11783 / J1939 \cite{J1939-81} defines address claiming for network entry, but lacks mechanisms for presence monitoring and graceful departure. ISO FastLane addresses this gap by repurposing two reserved Source Address values:

\begin{itemize}
\item \textbf{SA 253 - Presence/Heartbeat:} A CF may periodically send an address-claimed message with SA=253 to indicate it is still alive on the network. This serves as heartbeat for late joiners and avoids the bad practice of sending real address claims as periodic heartbeat.
\item \textbf{SA 252 - Departure:} Before shutting down, a CF shall send an address-claimed message with SA=252 to announce graceful departure. Upon receiving this, peers shall close any FastLane sessions and remove the CF from their address table.
\end{itemize}

This provides a complete device lifecycle: Entry (standard ACL), Presence (SA 253), and Exit (SA 252). Unexpected disconnections are handled via timeout detection. 

\section{The Messaging}
\label{subsec:techsolution}

\subsection{CAN Frames on Ethernet}

As the key for this solution is simplicity, we need to find also the most simple way to transfer J1939 frames on IP network. Here we assume that fundamental layers for IP communication are already in place, such as IP addresses and physical connectors etc. This is prerequisite for the ISO FastLane concept and out of the scope of this paper. 

As the ISO 11783 communication stack implementations have already a clear place where this transmission or reception of CAN frames happens (above HAL), this design must fit to this place without any other complexity. Our selection is to use UDP messages with fixed port number for reception. Our intention is to use registered port for this purpose. As we are logical, we will use IP port 11783 for this purpose, which is currently unassigned.

On transmission side, when any J1939 frame is sent, it must be checked whether it goes to physical CAN bus, or it goes to ISO FastLane. At this point, the available information is: 1) Priority, 2) Parameter Group Number (PGN), 3) Source Address (SA) and 4) variable length data of frame (0-8 bytes). If the data goes to CAN bus, information 1-3 is fused to CAN Identifier (29-bit) and information is put to CAN Transmission register with the data bytes. As we see some potential in following the idea of CAN frame, our proposal is to fuse information 1-3 also for ISO FastLane. Therefore the content of frame will be like in Table~\ref{tab:udp_format}.

The routing decision is simple: if the destination SA has FastLane enabled (known from AACL exchange), send via UDP; otherwise send via physical CAN bus. This is a Layer 3 routing decision that can be implemented as a simple if/else statement in existing ISOBUS stacks.

On reception side, the received UDP messages are processed with the same pipeline as the messages originating from physical CAN bus controller. Buffering of UDP message reception needs to be implemented in the same fashion as reception of CAN frames.

\subsection{Session Confirmation via 253-over-UDP}

After AACL exchange, both CFs have each other's IP addresses, but the UDP path may not be functional (firewall, NAT, etc.). To confirm the path works before sending application traffic, each CF shall send address-claimed messages with SA=253 encapsulated in UDP (253-over-UDP) to the peer.

Only after receiving 253-over-UDP from a peer shall a CF enable FastLane for that peer and begin routing point-to-point traffic via UDP. This ensures mutual confirmation that the UDP path is functional.

The 253-over-UDP also serves as session keepalive:
\begin{itemize}
\item \textbf{Period:} Each CF shall send 253-over-UDP to each FastLane peer at least every 5 seconds
\item \textbf{Timeout:} If no 253-over-UDP is received from a peer for 15 seconds, the FastLane session shall be considered failed and traffic shall fall back to CAN bus
\end{itemize}

This mechanism replaces the originally proposed Type 0 keep-alive message, providing both session confirmation and keepalive in one mechanism.

In the case of timeout, the application layer shall be informed and the same procedure shall take place as application specific timeouts (VT status message timeout, TC status message timeout, etc.). These defined procedures restart the session in a deterministic way, and for ISO FastLane this guarantees that communication falls back to CAN bus after Ethernet failure. 

\begin{table}[htbp]
\caption{ISO FastLane UDP Frame Format (Version 1)}
\label{tab:udp_format}
\centering
\begin{tabular}{|c|l|l|}
\hline
\textbf{Byte} & \textbf{Name} & \textbf{Description} \\
\hline
0 & Version & Protocol version. Value: 0x01 \\
\hline
1 & Type & Message type. Value: 0x01 (CAN frame) \\
\hline
2-5 & CAN ID & 32-bit CAN identifier (big-endian) \\
\hline
6 & DLC & Data Length Code (0-8) \\
\hline
7-14 & Data & CAN data bytes (always 8 bytes, padded with 0xFF) \\
\hline
\end{tabular}
\end{table}

The frame format is fixed at 15 bytes, allowing simple memcpy operations from existing CAN frame buffers. Padding with 0xFF follows J1939/ISO 11783 convention.

\textbf{Reserved for future:} Type 0x02 is reserved for CAN FD frames, which would extend the data field to support up to 64 bytes.

\textbf{No CRC required:} UDP provides its own checksum (mandatory in IPv6, optional but typically enabled in IPv4). Additionally, Ethernet includes CRC at the frame level. ISO FastLane requires wired networks; wireless is prohibited due to packet loss concerns.

\section{POC IMPLEMENTATION AND RESULTS}
\label{sec:implementation}
\subsection{Introduction}
\label{subsec:impl_intro}

To evaluate ISO FastLane performance on real hardware, we implemented the complete protocol stack for proof-of-concept (PoC), including CAN hardware abstraction, J1939 address claiming, Transport Protocol (TP) and Extended Transport Protocol (ETP), and the ISO FastLane routing layer and AACL. The implementation follows a single-threaded design with 5 ms tick interval. Two test scenarios were designed: a token-ring relay test measuring latency and throughput for multi-packet transfers, and a bidirectional stress test measuring maximum sustainable message rates. In both test scenarios, all programs are connected to real CAN bus segment at 250~kbit/s using USB-CAN adapters. In this PoC implementation, ISO FastLane UDP traffic used localhost (127.0.0.1) sockets on PC, simulating the Ethernet path with negligible network latency to isolate protocol-level performance from network effects. Each test was run with FastLane disabled (CAN only) and enabled (CAN + point-to-point via UDP), with identical application-level workloads.

\subsection{Token-Ring Relay Test}
\label{subsec:impl_tokenring}

The token-ring test uses three nodes (A, B, C) on separate CAN channels connected to the same physical bus. Node~A generates a random payload, sends it to Node~B via single-frame, TP or ETP. Node~B forwards the received data unmodified to Node~C, which forwards it back to Node~A. Node~A verifies byte-for-byte data integrity after the round trip. This true relay pattern exercises the complete protocol stack at each hop: address resolution, TP/ETP segmentation and reassembly with RTS/CTS flow control, and the FastLane routing decision.

The test matrix covers payload sizes from 8~bytes (single CAN frame) through 1~MB (Extended Transport Protocol), with CTS packet counts of 16 and 128 to evaluate flow control impact. The CTS value determines how many data packets the receiver requests per CTS handshake. Higher values reduce protocol overhead but increase burst sizes.

Table~\ref{tab:relay_results} presents the relay test results. With ISO FastLane enabled, all point-to-point TP/ETP traffic is routed via UDP, reducing CAN bus load to 0\% for data transfer. The speedup factor depends strongly on the CTS packet count: at CTS=16 (the ISO~11783 \cite{ISO11783-3} recommended default), frequent handshake round-trips on the CAN bus limit throughput to approximately 32-37\% bus utilization, yielding 2.4-3.7x speedup. At CTS=128, the CAN bus reaches 74-76\% utilization before FastLane activation, and the speedup increases to 5.3-7.9x.

\begin{table}[htbp]
\caption{Token-ring relay results (3 nodes, A$\to$B$\to$C$\to$A)}
\label{tab:relay_results}
\centering
\small
\begin{tabular}{|l|r|r|r|r|}
\hline
\textbf{Payload} & \textbf{CAN} & \textbf{FL} & \textbf{Speed-} & \textbf{CAN} \\
\textbf{(CTS)} & \textbf{ms/rnd} & \textbf{ms/rnd} & \textbf{up} & \textbf{bus\%} \\
\hline
8B single frame   & 33    & 14    & 2.5x  & 14.7  \\
\hline
1000B (CTS=16)    & 679   & 182   & 3.7x  & 36.7  \\
1000B (CTS=128)   & 371   & 70    & 5.3x  & 64.2  \\
\hline
1785B (CTS=16)    & 1208  & 328   & 3.7x  & 36.6  \\
1785B (CTS=128)   & 563   & 71    & 7.9x  & 74.4  \\
\hline
5KB (CTS=16)      & 4144  & 1685  & 2.5x  & 31.5  \\
5KB (CTS=128)     & 1660  & 257   & 6.5x  & 71.1  \\
\hline
50KB (CTS=128)    & 15543 & 1964  & 7.9x  & 75.6  \\
\hline
1MB (CTS=16)      & 826k & 343k & 2.4x  & 31.5  \\
1MB (CTS=128)     & 312k & 42k  & 7.5x  & 75.4  \\
\hline
\end{tabular}
\end{table}

The 1~MB relay test (5 rounds $\times$ 3 hops = 15~MB total verified data) demonstrates the practical impact: a round trip that takes 5.2~minutes on CAN bus completes in 42~seconds with FastLane (CTS=128). This corresponds to real-world scenarios such as VT object pool uploads or large File Server data transfers. Zero data errors were observed across all test configurations. Every byte of the 15~MB+ total transferred data was verified.

The remaining 25\% gap to theoretical CAN bus saturation at CTS=128 is attributable to CTS/DPO handshake overhead and inter-frame gaps, which are inherent to the TP/ETP protocol regardless of physical transport.

\subsection{Bidirectional High-Rate Stress Test}
\label{subsec:test2_turbo}

To evaluate FastLane performance under conditions exceeding CAN bus capacity, a bidirectional stress test was designed. Two nodes (A and B) simultaneously send point-to-point frames to each other at escalating rates. At 250~kbit/s with approximately 135~bits per extended CAN frame (depending on bit stuffing), the theoretical CAN bus maximum is at maximum about 2000~frames/sec. Any bidirectional rate above 1000~frames/sec per node would saturate the bus.

Each frame carries an 8-byte payload: 1~byte sender identifier, 4~bytes sequence number, and 3~bytes random data for payload diversity. Frames are generated at a fixed rate per 5~ms tick (e.g., 10~frames/tick = 2000~frames/sec per node). The test runs for 10~seconds at each rate, with a 500~ms grace period for in-flight frames. Loss is calculated per direction by comparing sent and received sequence counts.

The target rate of 2000~msg/s per node (4000~bidirectional) represents a 10x improvement over the current ISO~11783-10 TC limit of (10~msg/s average per message type etc. rules). This would enable significantly faster section control and multirate control for modern planters and sprayers. We also explored the upper throughput limit to characterize system capacity.

Table~\ref{tab:throughput_results} summarizes the results. The CAN-only baseline at 200~msg/s shows occasional frame loss (0.1\%) from bus contention, while FastLane at the same rate achieves zero loss. With FastLane enabled, all rates from 200 through 80000~msg/s per node were sustained with zero packet loss: a range spanning from 22\% to 8640\% of CAN bus equivalent capacity.

\begin{table*}[!t]
\caption{Bidirectional throughput stress test results (2 nodes, simultaneous send)}
\label{tab:throughput_results}
\centering
\small
\begin{tabular}{|r|r|l|r|r|r|r|l|}
\hline
\textbf{Rate} & \textbf{Per} & \textbf{Mode} & \textbf{Sent} & \textbf{Rcvd} & \textbf{Loss} & \textbf{Equiv. CAN} & \textbf{Result} \\
\textbf{(msg/s/node)} & \textbf{tick} & & \textbf{per dir.} & \textbf{per dir.} & & \textbf{bus load} & \\
\hline
200    & 1     & CAN       & 2000     & 1998     & 0.10\%    & 21.6\%     & PASS \\
200    & 1     & FastLane  & 2000     & 2000     & 0.00\%    & 21.6\%     & PASS \\
1000   & 5     & FastLane  & 10000    & 10000    & 0.00\%    & 108\%      & PASS \\
2000   & 10    & FastLane  & 20000    & 20000    & 0.00\%    & 216\%      & PASS \\
4000   & 20    & FastLane  & 40000    & 40000    & 0.00\%    & 432\%      & PASS \\
10000  & 50    & FastLane  & 100000   & 100000   & 0.00\%    & 1080\%     & PASS \\
20000  & 100   & FastLane  & 200000   & 200000   & 0.00\%    & 2160\%     & PASS \\
40000  & 200   & FastLane  & 400000   & 400000   & 0.00\%    & 4320\%     & PASS \\
80000  & 400   & FastLane  & 800000   & 800000   & 0.00\%    & 8640\%     & PASS \\
90000  & 450   & FastLane  & 900000   & 887998   & 1.33\%    & 9720\%     & FAIL \\
\hline
\end{tabular}
\end{table*}

First packet loss occurred at 90000~msg/s per node (1.33\%), with both nodes receiving exactly 887998~frames, indicating the UDP socket receive buffer (1~MB) as the bottleneck rather than CPU or network capacity. The maximum lossless bidirectional throughput of 160000~msg/s (80000 per node) represents 86 times the CAN bus capacity. A sustained 30-second test at 2000~msg/s per node (120000~frames total per direction) confirmed zero loss over extended operation.

For practical ISO~11783 applications, even the conservative 2000~msg/s rate provides a 10x improvement for TC section control and rate control, while the demonstrated 80000~msg/s headroom leaves ample margin for future application requirements. On our localhost PoC, the zero-loss reliability across all tested rates below the system limit confirms that ISO FastLane provides deterministic communication suitable for real-time agricultural control.

\section{Discussion}
\label{sec:discussion}

\subsection{Benefits}
\label{subsec:benefits}

The proposed solution is for short-term, but lifetime of short-term solution is unknown as it will depend on market adaptation. This proposal is based on idea that in the near future every tractor and implement will still have IBBC connector, which provides both 12V/55A power and CAN bus based ISO 11783. As the long-term solution aims at functionalities which work without CAN bus, this short-term solution will fade away along with IBBC connector, one day when long-term solution is on market. This is quite similar future scenario as ISO 11786 signal connector vs. ISO 11783 bus: if ISO 11783 had quality issues, farmers continued using ISO 11786 until technology was ready for them. The same will happen in the future with long-term and short-term solution with HSI.

The main benefit of this ISO FastLane proposal is simplicity and simple rules. It does not take into account how the basic network is realized, as long as two CF's will get their counterparty IP address and UDP messages flow through the network without firewall. The short-term solution requires also from AEF simple basic (IP) network, which is needed even for Digital Camera Systems.

The network efficiency of ISO FastLane is not optimized, as using one UDP packet for each CAN frame results in protocol overhead. Each 8-byte CAN payload is encapsulated in a 15-byte FastLane frame, plus UDP (8 bytes), IP (20 bytes), and Ethernet (14 bytes + 4 byte FCS) headers, approximately 61 bytes total per CAN frame. However, our measurements demonstrate this overhead is negligible in practice: the bidirectional stress test sustained 80000~msg/s per node with zero loss, achieving approximately 0.64 Megabytes/s of payload throughput. The available Ethernet bandwidth (100~Mbit/s or 1~Gbit/s) vastly exceeds the original CAN capacity (250~kbit/s), making protocol efficiency a non-issue for this short-term solution.

The practical benefits of ISO FastLane span all three major application layers. For Virtual Terminals, faster object pool uploads reduce startup time, our 1~MB ETP relay completed in 42~seconds via FastLane versus 5.2~minutes on CAN (CTS=128), a 7.5x improvement. Smaller pool transfers (1785~bytes, typical for incremental updates) showed 7.9x speedup (71~ms vs 563~ms). These improvements directly translate to reduced operator wait times when switching implements or restarting systems during field operations.

For Task Controllers, the increased message throughput enables higher-rate automation control loops. Where CAN-constrained implementations were limited to approximately 10 messages per second average, our stress test demonstrated sustained 2000~msg/s with zero loss, which is a 200x improvement. Even at 80000~msg/s per node, no packets were lost. This enables more precise section control, faster valve adjustments, and smoother implement behavior during high-speed field operations.

File Server applications benefit most dramatically from the bandwidth improvement. Large data files such as field maps, prescription data, or firmware updates that previously required extended transfer windows can now be transmitted during normal operation without significantly impacting other CAN bus traffic. The 100\% reduction in CAN bus load for point-to-point traffic (confirmed across all relay test configurations) also benefits legacy devices sharing the bus, as broadcast messages experience zero contention from FastLane traffic.

\subsection{Backward Compatibility}
\label{subsec:backwards_compat}

ISO FastLane maintains full backward compatibility with legacy ISO 11783 devices:

\textbf{Legacy Device Transparency:} Legacy devices without Ethernet capability continue to operate normally on CAN bus. They are unaware of FastLane traffic occurring on Ethernet between other devices.

\textbf{Graceful Degradation:} ISO FastLane-capable devices detect legacy peers (those that do not respond to AACL requests) and communicate with them exclusively via CAN bus. No special configuration is required.

\textbf{Mixed Networks:} A VT server with FastLane can simultaneously serve FastLane-capable clients via Ethernet and legacy clients via CAN bus. The routing decision is per-peer.

\textbf{No Protocol Changes:} The message content (PGN, data fields, timing semantics) remains unchanged. Message throttling limits can be disabled for ISO FastLane. Only the transport layer differs. Application layer code is unaffected.

\textbf{Fallback Mechanism:} If Ethernet communication fails (detected via 253-over-UDP timeout), the session falls back to CAN bus using standard ISO 11783 timeout and restart procedures.

\subsection{Security Considerations}
\label{subsec:security}

ISO FastLane inherits the security model of ISO 11783, which was designed for closed agricultural networks without encryption or authentication. The following security aspects apply:

\textbf{Network Isolation:} The HSI network should be isolated from external networks. No internet gateway should be present on the vehicle network segment used for ISO FastLane.

\textbf{No Encryption:} UDP traffic is unencrypted, matching the unencrypted nature of CAN bus. Adding encryption would increase complexity and latency, contradicting short-term solution goals.

\textbf{Trust Model:} All devices on the network are implicitly trusted, as with current ISO 11783. Physical access control to connectors is the primary security boundary.

\textbf{Spoofing Risk:} An attacker with network access could spoof UDP packets. This is equivalent to the existing risk of CAN message injection. Long-term solutions should address authentication.

For high-security applications, ISO FastLane should be deployed only on physically secured networks. Security enhancements are deferred to the long-term solution which may incorporate modern security mechanisms. On the other hand, security can be also handled on higher layers, e.g. Tractor-Implement Management (TIM) works over ISO FastLane secured, similarly as over CAN bus. 

\subsection{Implications for J1939 Standardization}
\label{subsec:standardization}

Standardizing ISO FastLane requires coordination with SAE J1939 and ISO 11783 committees:

\textbf{PGN Assignment:} The AACL message requires a formal PGN assignment from SAE. For this prototype, PF=75 (PGN 19200) is used as a placeholder. The final PF value must be determined through SAE's formal standardization process and documented in SAE J1939/71 \cite{J1939-71} or ISO 11783-5 \cite{ISO11783-5}.

\textbf{Reserved SA Usage:} The use of SA value 252 (departure) and SA value 253 (presence) repurposes currently reserved values (not real addresses). Similarly, already SA value 254 is already having special meaning. This requires adoption in SAE J1939/81 \cite{J1939-81} (Network Management) and/or ISO 11783-5 \cite{ISO11783-5}.

\textbf{ISO 11783-2 Amendment:} The physical layer standard \cite{ISO11783-2} may need amendment to formally recognize HSI as a secondary interface option alongside the existing CAN bus requirement.

\textbf{Application Layer Amendments:} To enable higher message rates on FastLane, amendments to ISO 11783-6 \cite{ISO11783-6} (VT), ISO 11783-10 \cite{ISO11783-10} (TC), and ISO 11783-13 \cite{ISO11783-13} (FS) may relax timing constraints when operating over Ethernet.

The standardization path could be either through ISO TC 23/SC 19 for agricultural applications, or through SAE J1939 for broader adoption across vehicle industries.

\subsection{Applicability to NMEA 2000}
\label{subsec:applicability_nmea}

NMEA 2000 \cite{NMEA2000}, used in marine electronics, is based on SAE J1939 \cite{J1939} and shares the same CAN bus architecture and similar constraints as ISO 11783 and SAE J1939. The ISO FastLane concept could be adapted for marine applications. However, NMEA 2000 is mainly built on broadcast messages, at least when it comes to GNSS related data. 

\subsection{Accessibility for J1939/ISO 11783 Engineers}
\label{subsec:implementation_accessibility}

A key design goal of ISO FastLane is accessibility for engineers already familiar with J1939/ISO 11783 stacks. The implementation requires minimal changes to existing code:

\textbf{Routing Decision:} The core change is a simple routing decision at OSI Layer 3. When transmitting a destination-specific message, the stack checks if the destination SA has FastLane enabled. This can be implemented as:

\begin{verbatim}
if (has_fastlane(destination_sa)) {
    send_via_udp(peer_ip, peer_port, can_frame);
} else {
    send_via_can(can_frame);
}
\end{verbatim}

\textbf{Address Table Extension:} The existing J1939 address table, which maps SA to NAME, is extended to include IP address and UDP port for FastLane-capable peers.

\textbf{Frame Format Preservation:} The UDP payload contains a standard CAN frame structure. Existing frame parsing code can be reused with minimal modification. The frame simply arrives from a different source (UDP socket vs CAN controller).

\textbf{No New Complex State Machines:} FastLane leverages existing J1939 mechanisms (address claiming, timeout handling) rather than introducing new state machines. The 253-over-UDP keepalive is the only new periodic task required.

This approach allows incremental adoption: engineers can add FastLane support to existing stacks without major architectural changes.                                                                 

\subsection{Startup Easiness}
\label{subsec:startup}

The startup sequence for ISO FastLane devices follows the established ISO 11783 procedures with minimal additions:

\begin{enumerate}
\item \textbf{Network Initialization:} ECU initializes both CAN and Ethernet interfaces. IP address must be available before proceeding (via DHCP, static configuration, or link-local).
\item \textbf{Address Claiming:} CF performs standard J1939 address claiming on CAN bus.
\item \textbf{Address Validity:} After 250ms without contention, the claimed address is valid.
\item \textbf{AACL Exchange:} CF requests AACL from other CFs on the bus. Responses received within timeout indicate FastLane-capable peers.
\item \textbf{Session Confirmation:} CF begins sending 253-over-UDP to peers with known IP addresses.
\item \textbf{FastLane Activation:} Upon receiving 253-over-UDP from a peer, FastLane is enabled for that peer.
\item \textbf{Normal Operation:} Point-to-point traffic routes via UDP; broadcast remains on CAN.
\end{enumerate}

If a CF's IP address changes during operation, it must restart the sequence from step 2 (re-claim address) to trigger AACL re-exchange with all peers.

\subsection{Performance Optimization in Practice}
\label{subsec:performance}

While ISO FastLane enables significantly faster communication, realizing its full potential may require attention to existing ISO 11783 stack implementations. When application layer throttle limits are removed (e.g., 10 msg/s for TC), all communication layers operate at higher speeds simultaneously. This acceleration can expose design compromises in existing stacks that were originally tailored for CAN's maximum throughput of approximately 2000 msg/s.

Common issues that may surface include hardcoded timer values based on CAN timing assumptions, fixed buffer sizes optimized for lower data rates, and processing bottlenecks that were previously masked by the slower CAN bus. For example, Extended Transport Protocol (ETP) performance depends critically on reaction times between packets. If internal processing is throttled by fixed timer intervals, the theoretical speed improvement from FastLane cannot be fully realized.

Additionally, the higher message rates enabled by FastLane make packet ordering more susceptible to disruption compared to the inherently serialized nature of CAN reception. Without careful synchronization, out-of-order delivery becomes more likely when messages arrive via both CAN and UDP simultaneously.

The implementation of ISO FastLane itself is straightforward when the original CAN-based ISOBUS stack follows a well-layered architecture with clear separation between hardware abstraction, network management, transport protocol, and application layers. The routing decision can be inserted cleanly at the network layer without affecting higher layers.

In practice, integrating ISO FastLane is likely to reveal latent bugs and performance limitations that should have been addressed regardless of FastLane adoption. Therefore, it is advisable to allocate sufficient resources for general quality improvements to the ISOBUS communication stack when planning an upgrade project to include ISO FastLane support.

\subsection{Proof-of-Concept Performance}
\label{subsec:poc_results}

The Proof-of-Concept implementation and preliminary experimental evaluation on physical CAN hardware and localhost sockets demonstrated significant performance improvements:

\begin{itemize}
\item Up to 7.9x faster TP/ETP transfers (563~ms $\rightarrow$ 71~ms for 1785~byte relay), with 1~MB round-trip reduced from 5.2~minutes to 42~seconds
\item 100\% CAN bus load reduction for point-to-point traffic, freeing bandwidth for broadcast messages
\item 86x CAN bus capacity demonstrated: 80000~msg/s per node sustained bidirectionally with zero packet loss
\item Zero data errors across 15~MB+ of verified relay data and 1.6~million stress test frames
\end{itemize}

Even if these preliminary results are achieved with localhost sockets, the numbers show the potential of acceleration. 

In addition, it should be noted, that in this implementation, the basic communication stack was not modified to accelerate timings of computing, but keeping normal 5 ms tick as used in CAN only processing. With complete redesign of ISO 11783 stack the acceleration could be much faster. The scope of this PoC implementation was to check functionality of ISO FastLane and see acceleration with as minimal changes as possible to the stack. 

\subsection{Limitations}
\label{subsec:limitations}

ISO FastLane has several intentional limitations that result from prioritizing simplicity over optimality:

\textbf{Protocol Overhead:} Each CAN frame (up to 8 data bytes) is encapsulated in a 15-byte UDP payload, plus UDP/IP/Ethernet headers. This results in approximately 60 bytes per message on Ethernet compared to the original CAN frame. While inefficient compared to optimized protocols, the available Ethernet bandwidth (1 Gbit/s vs 250 kbit/s) makes this overhead negligible in practice.

\textbf{No Multicast or Broadcast on Ethernet:} All FastLane communication is unicast UDP. Broadcast messages remain on CAN bus exclusively. This is intentional to avoid complexity with managed switches and multicast group management.

\textbf{Wired Networks Only:} Wireless networks are explicitly prohibited due to packet loss affecting real-time control. This limits deployment flexibility but ensures deterministic behavior.

\textbf{Single Point of Failure:} If Ethernet connectivity fails, the system falls back to CAN bus. However, if both fail simultaneously, no communication is possible. This is inherent to any dual-stack approach.

\textbf{No Quality of Service:} FastLane does not implement QoS mechanisms on Ethernet. In simple point-to-point networks this is acceptable, but complex networks with many devices may require switch configuration for traffic prioritization.

\textbf{Future High Speed Isobus:} ISO FastLane is intended as the short-term solution. A long-term solution should not rely on CAN frames or J1939 messaging. ISO FastLane is limited to dual stack principle, which assumes that ECU has physical CAN bus interface either directly built-in to ECU, or indirectly via proprietary gateway of same OEM (see: fundamental idea A). 

\subsection{Future Work}
\label{subsec:future_work}

Several extensions to ISO FastLane are envisioned for future development:

\textbf{Formal Standardization:} While PGN 19200 has been assigned for AACL, formal inclusion in SAE J1939/71 \cite{J1939-71} (Application Layer) or ISO 11783-5 \cite{ISO11783-5} (Network Management) standards is recommended to ensure interoperability across implementations.

\textbf{Application Layer Rate Limit Relaxation:} Current ISO 11783 application layers impose rate limits designed for CAN bus constraints (e.g., TC average 10 messages per second). To fully exploit FastLane bandwidth, these limits should be relaxed for sessions operating over Ethernet. This requires amendments to ISO 11783-6 \cite{ISO11783-6} (VT), ISO 11783-10 \cite{ISO11783-10} (TC), and ISO 11783-13 \cite{ISO11783-13} (FS).

\textbf{CAN FD Support:} The UDP frame format reserves Type 0x02 for CAN FD frames, which would extend the data field from 8 bytes to support up to 64 bytes. This would enable encapsulation of CAN FD traffic over Ethernet while maintaining the same simple routing architecture. CAN FD is already specified in ISO 11783-2 \cite{ISO11783-2} as a future option.

\textbf{Wireless Considerations:} ISO FastLane explicitly prohibits wireless networks due to packet loss concerns affecting real-time control. However, future work could explore reliable wireless protocols with appropriate error correction for non-critical data transfer, such as pool uploads or file transfers.

\section{Conclusion}
\label{sec:conclusion}

This paper presented ISO FastLane, a dual-stack approach for accelerating ISO 11783 communication using Ethernet alongside the mandatory CAN bus. 

The benefits of short-term solution over long-term solution are always limited, as time to market is more important than the optimal solution. Secondarily, standardized technology should be easy to implement for everyone without complex rules. Usually complex rules have led to different interpretations by different companies and furthermore this has led to complex conformance test tools and expenses and delays related to this. Gateway-less ISO FastLane is a perfect short term solution for fast ISOBUS systems, intended for more dynamic Virtual Terminal, fast rate Task Controller and large file File Server.

The implementation requires minimal changes to existing ISO 11783 stacks: a simple Layer 3 routing decision, AACL message handling, and UDP socket management. Application layer code remains unchanged, demonstrating that ISO FastLane can be incrementally adopted without major architectural modifications. 

ISO FastLane provides short-term solution especially for high-tech Task Controller applications, which currently are not possible for multirow planters 

\section*{Acknowledgements}
\label{sec:acknowledgements}

The initial idea to speed up ISOBUS over Ethernet using secondary channel with a metaphfora "hey buddy, we both have much faster secondary channel, let's chat more over there" was proposed by Ludger Autermann in November 2024. This "gateway-less" architecture was also part of his initial idea, for simplicity. With his consent, the author derived the initial idea further independently, to extend the idea to cover all presented fundamental ideas and to develop and engineer all details to get from the initial rough idea to this functional and proven concept called ISO FastLane. The author wants to thank Ludger Autermann for this inspiring and clever initial idea, which motivated to do the further engineering and complete this work for a solid and comprehensive standardization proposal. 

The author wants to thank also VDMA Landtechnik for long-term support for our research, since 2019.

\bibliographystyle{IEEEtran}
\bibliography{literature}

\end{document}